\documentclass[aps,%
reprint,
superscriptaddress,
amsmath,amssymb,floatfix,pra,
]{revtex4-2}
\usepackage{lipsum}
\usepackage{graphicx}% Include figure files
\usepackage{dcolumn}% Align table columns on decimal point
\usepackage{comment}
\usepackage{bm}% bold math
\usepackage{color}
\usepackage{graphicx}% Include figure files
\usepackage{dcolumn}% Align table columns on decimal point
\usepackage{comment}
\usepackage{bm}% bold math
\usepackage{xcolor}
\definecolor{darkblue}{rgb}{0,0,0.6}
\usepackage{hyperref}
\hypersetup{colorlinks,linkcolor=darkblue,citecolor=darkblue,urlcolor=darkblue}

\usepackage[normalem]{ulem}

\begin{document}

\title{Hyperuniform interfaces in non-equilibrium phase coexistence}

\author{Rapha\"el Maire}
%\email{maire.raphael@yahoo.fr}
\affiliation{Universit\'e Paris-Saclay, CNRS, Laboratoire de Physique des Solides, 91405 Orsay, France}

\author{Leonardo Galliano}

\affiliation{Dipartimento di Fisica, Universit\`a di Trieste, Strada Costiera 11, 34151, Trieste, Italy}

\affiliation{Gulliver, CNRS UMR 7083, ESPCI Paris, PSL Research University, 75005 Paris, France}

\author{Andrea Plati}

\affiliation{Universit\'e Paris-Saclay, CNRS, Laboratoire de Physique des Solides, 91405 Orsay, France}

\author{Ludovic Berthier}

\affiliation{Gulliver, CNRS UMR 7083, ESPCI Paris, PSL Research University, 75005 Paris, France}

\date{\today}

\begin{abstract}
We show that long-wavelength interfacial fluctuations are strongly suppressed in non-equilibrium phase coexistence between bulk hyperuniform systems. Using simulations of three distinct microscopic models, we demonstrate that hyperuniform interfaces are much smoother than equilibrium ones, with a universal reduction of height fluctuations at large scale. We derive a non-equilibrium interface equation from the field theory of the bulk order parameter, and predict a reduction in height fluctuations, $S_h(\bm k)\equiv \langle |h(\bm k)|^2\rangle\sim |\bm k|^{-1}$, in stark contrast to equilibrium capillary wave theory where $S_h(\bm k)\sim |\bm k|^{-2}$. Our results establish a new universality class for non-equilibrium interfaces, highlighting the fundamental role of suppressed bulk fluctuations in shaping interfacial dynamics far from equilibrium.
\end{abstract}

\maketitle

Interfaces and phase separations are fundamental phenomena driving the organization and dynamics of matter across scales, from soft condensed matter systems~\cite{jones2002soft, onuki2002phase, zaccarelli2007colloidal, roos2015phase} to complex biological assemblies~\cite{hyman2014liquid,jacobs2017phase, sinhuber2017phase, alberti2017phase, feigenson2007phase} or granular systems~\cite{plati2024self, maire2024non, clewett2016minimization, roeller2011liquid, mujica2016dynamics}. These processes regulate the behavior of systems where distinct phases coexist and interact, often leading to intricate interfacial dynamics~\cite{barabasi1995fractal} possibly shaping the geometry and controlling the physics of complex materials. 

An equilibrium interface in a phase‑separated system obeys capillary wave theory, with a Hamiltonian on large scales given by the excess area of the fluctuating interface~\cite{onuki2002phase}: $\mathcal{H}[h]= \gamma\bm k^2 h({\bm k})/2$, where $h({\bm k})$ is the Fourier transform of the interface height and $\gamma$ the surface tension. By equipartition, the resulting static height correlations obey $S_h(\bm k)\equiv\langle |h(\bm k)|^2\rangle=k_BT\bm |\bm k|^{-2}/\gamma$, with $T$ the temperature, and diverges as $|\bm k|^{-2}$, a signature of massless Goldstone modes associated with broken continuous translational symmetry~\cite{zia1988dynamics}, in analogy with vibrational dynamics in crystals~\cite{PhysRevLett.131.047101}. 

In contrast, non-equilibrium systems can exhibit qualitatively different interfacial behavior. While growing interfaces--such as those described by the Kardar-Parisi-Zhang equation--are well-studied examples~\cite{barabasi1995fractal}, we focus instead on phase-separated systems with stable, non-growing interfaces. In this case, the interface is driven by the non-equilibrium nature of the bulk phases that coexist. This can arise through external driving, such as boundary forcing~\cite{grant1983fluctuating, nakagawa2017liquid, sasa2025non} or shear~\cite{bray2001interface, derks2006suppression}, leading to phenomena including traveling waves~\cite{barthelet1998benjamin}, reduced surface tension~\cite{derks2006suppression, del2019interface}, and interfacial instabilities~\cite{onuki2002phase}.

Internally driven systems--such as active matter--offer further richness. These can sustain novel interfacial dynamics, including persistent traveling modes~\cite{adkins2022dynamics, langford2024phase, rasshofer2025capillary} and instabilities~\cite{cates2024active, zhao2024asymmetric}. A vast majority of experimental and numerical investigations of stable interfaces in active and granular matter report $S_{h}(\bm k)\sim |\bm k|^{-2}$, in agreement with capillary wave theory, albeit with an effective surface tension and temperature~\cite{luu2013capillarylike, yao2025interfacial, Wysocki_2016, langford2024theory, patch2018curvature, lee2017interface, paliwal2017non} that account for the deviations from equilibrium. Recently, scalar active interfaces have been theoretically predicted to exhibit scaling laws distinct from equilibrium capillary behavior~\cite{besse2023interface, caballero2024interface}, with some numerical support~\cite{krishan2025finite}.

Unlike many active systems, whose bulk and interfacial large-scale properties often resemble equilibrium behavior~\cite{fodor2022irreversibility}, non-equilibrium hyperuniform systems constitute a distinct class of matter characterized by anomalous fluctuations that defy equilibrium expectations~\cite{torquato2018hyperuniform,lei2024non}. Found in diverse contexts such as optimization problems in nature~\cite{PhysRevE.89.022721, atkinson2016critical, Ge_2023}, driven colloids~\cite{tjhung2015hyperuniform} or cosmology~\cite{philcox2023disordered}, these systems are of wide interest for their potential to form materials with novel physical properties~\cite{man2013isotropic, shi2025three}. Hyperuniformity is defined by the suppression of long-wavelength bulk density fluctuations $\delta \rho({\bm k})$, with a vanishing bulk structure factor, $S_\rho(\bm k) \equiv \langle | \delta \rho(\bm k) |^2 \rangle$, at small $\bm k$~\cite{torquato2003local}. Such behavior is forbidden in equilibrium systems with short-range interactions at finite temperature, where instead $S_\rho({\bm k})$ converges to a finite value~\cite{torquato2018hyperuniform}.

Here we study the statistical properties of the interface in phase coexistence between non-equilibrium hyperuniform systems. We construct and analyze three distinct microscopic models and observe that bulk hyperuniformity is inherited by the interface, which displays suppressed height fluctuations at large scale. We derive an interface equation that explains the observed behavior, which can be summarized as:  
\begin{equation*}
\begingroup
\setlength{\arraycolsep}{9pt}
\renewcommand{\arraystretch}{1.5}
\begin{array}{c|@{\quad}c@{\quad}c@{\quad}c}
t\to\infty \text{ and } L\text{ large}& W^2_{1d} & W^2_{2d} & S_h(\bm{k}) \\
\hline
\text{Equilibrium} & \sim L & \sim \log L & \sim |\bm{k}|^{-2} \\
\text{Hyperuniform} & \sim \log L & \text{const.} & \sim |\bm{k}|^{-1}
\end{array}
\endgroup
\end{equation*}
with $ W^2(t)\equiv\left\langle[\langle h(\bm r, t)\rangle - h(\bm r, t)]^2\right\rangle$, the interface squared mean width. More generally, scale-free interfaces are characterized by two critical exponents which can be measured by starting from a flat interface at $t=0$ in systems of linear size $L$~\cite{onuki2002phase}:
\begin{equation}
W^2(t) \sim 
\begin{cases}
L^{2\chi} & \text{if } t \to \infty \\
t^{2\chi/z} & \text{if } L \to \infty
\end{cases}, \quad
S_h(\bm{k}) \sim |\bm{k}|^{-d - 2\chi}.
\end{equation}
The dynamical exponent $z$ is equivalently obtained from the relaxation time to the steady state, $\tau_{\rm ss}(L) \sim L^z$. Hyperuniform interfaces then display a suppression of height fluctuation from $|{\bm k}|^{-2}$ to $|{\bm k}|^{-1}$, and a reduction of the roughness exponent $\chi$ from $(2-d)/2$ to $(1-d)/2$. 

\begin{figure*}[!ht]
\includegraphics[width=0.98\textwidth,clip=true]{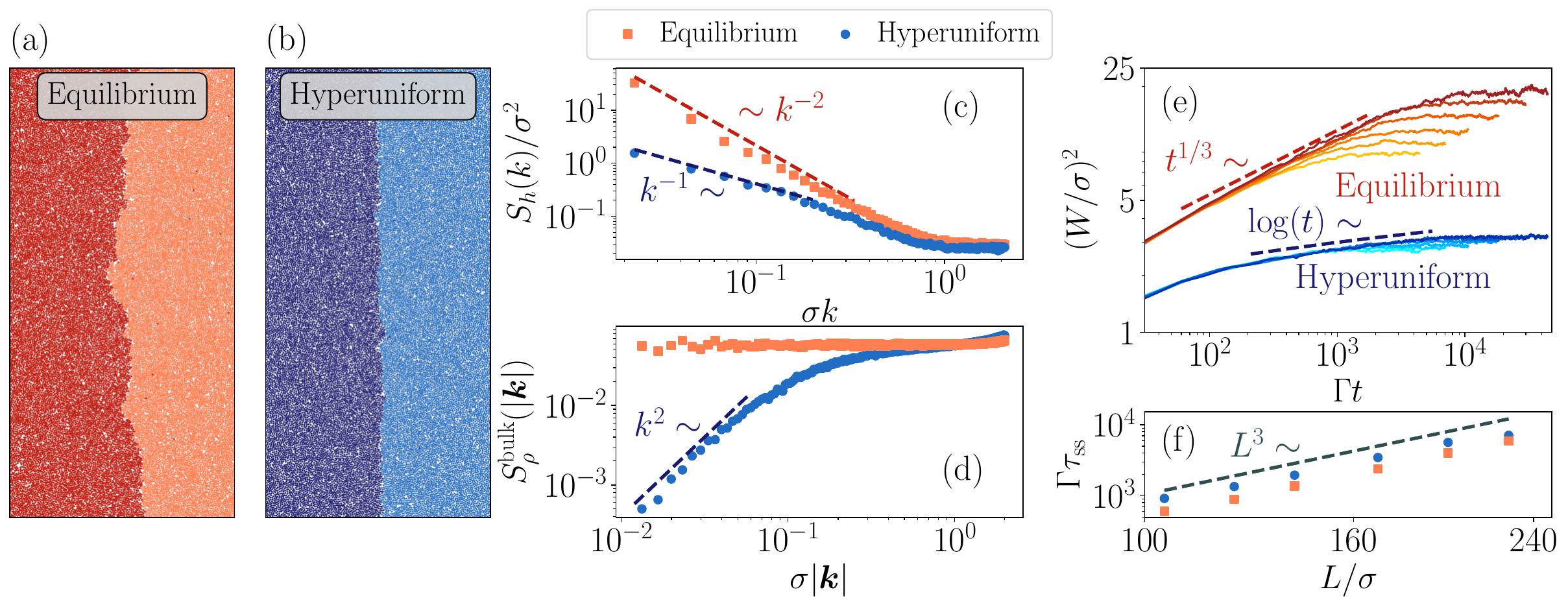} 
\centering
\caption{Binary-mixture with active collisions. 
(a) Snapshot of a fully phase-separated $AB$ mixture evolving under equilibrium Langevin dynamics with a square well potential interaction. The parameters are:  $T/U=-0.8$, $\sigma_U/\sigma=1.6$, $m(\sigma\Gamma)^2/T=0.075$, $N\pi\sigma^2/(4L_xL_y)=0.55$ and $L_x=1.81L_y$, where $T$ is the kinetic energy of the system per particle. 
(b) Snapshot for the non-equilibrium system governed by Eqs.~(\ref{eq: collision rule RanNi}, \ref{eq: damping}) with attraction. All parameters are identical to those in the equilibrium case without thermostat, and $\Delta E$ is chosen to match the kinetic energy of the equilibrium system. 
(c) Interface static height correlation functions for the two cases. 
(d) Radially averaged structure factors in the bulk. 
(e) Time evolution of $W^2$ for various system size for both models starting from a flat initial interface, averaged between 80 and 900 independent realizations. 
(f) Time to reach steady state $\tau_{\rm ss}$ as a function of the short box length $L \equiv L_y$ for the systems in (e).}   
\label{fig: comparison}
\end{figure*}

We have numerically measured these reduced interface fluctuations for three distinct hyperuniform phase separated models with one-dimensional interfaces. A first mechanism to produce bulk hyperuniformity is to locally inject energy through non-equilibrium momentum-conserving collisions combined with global dissipation via viscous damping~\cite{lei2024non}. This provides a robust framework for the spontaneous emergence of bulk hyperuniformity in non-equilibrium fluids~\cite{maire2025hyperuniformity}. We consider a system of hard disks of mass $m$ and diameter $\sigma$. When two particles $i$ and $j$ collide, they undergo a momentum-conserving collision accompanied by a constant energy increment $\Delta E>0$:
\begin{equation}
    \begin{split}
    \bm v_i'+\bm v_j'&=\bm v_i+\bm v_j,\\
    \dfrac{1}{2}m\left( {\bm v_i'}^2+{\bm v_j'}^2\right)&=\dfrac{1}{2}m\left(\bm v_i^2+\bm v_j^2\right)+\Delta E,
    \end{split}
    \label{eq: collision rule RanNi}
\end{equation}
where $\bm v'_i$ and $\bm v_i$ are the post- and pre-collisional velocities of the particle $i$, respectively. Particles are subject to a viscous damping $\Gamma$ during their free flight:
\begin{equation}
    \bm v(t) = \bm v(0)e^{-\Gamma t}.
     \label{eq: damping}
\end{equation}
In the model defined by Eqs.~(\ref{eq: collision rule RanNi}, \ref{eq: damping}), phonons cannot propagate and the system becomes hyperuniform~\cite{enhancing2024Maire, lei2019hydrodynamics}. 

To induce phase separation and create a non-equilibrium interface, we introduce attractive interactions in addition to the non-equilibrium energy injecting collisions. We use a binary mixture with particles of types $A$ and $B$, with attractive $AA$ and $BB$ interactions, but no attraction for $AB$. As in equilibrium, this results in a liquid-liquid phase separation at large enough attraction~\cite{oku2020phase}. The attraction is modeled as a square potential with depth $U$ and width $\sigma_U$. This model system can be regarded an idealization of a weakly cohesive quasi-2d vibrated granular gas. We will compare this system to an equilibrium one in which collisions are elastic, $\Delta E=0$, and an uncorrelated Gaussian white noise with variance $2\Gamma T$ is added to thermalize the system.

We perform event-driven molecular dynamics simulations~\cite{smallenburg2022efficient} of both equilibrium and non-equilibrium systems in a two-dimensional system of $N$ particles in an elongated periodic box with $L_y<L_x$, using a 50:50 ratio of $A$ and $B$ particles. We operate away from the critical point, deep in the phase coexistence region. As shown in Figs.~\ref{fig: comparison}(a, b), the interface separating the two phases in the equilibrium system appears rougher than in the non-equilibrium system, despite the bulk phases in both cases being tuned to have the same average kinetic energy. This visual difference is quantified by the static height correlation $S_h$ in Fig.~\ref{fig: comparison}(c). The equilibrium system exhibits the expected $|\bm k|^{-2}$ divergence, characteristic of thermal fluctuations. Instead, the interface separating the hyperuniform bulk fluids shows strongly suppressed fluctuations at large scale, compatible with a smaller $|\bm k|^{-1}$ divergence. We confirm in Fig.~\ref{fig: comparison}(d) that the bulk of each phase in the non-equilibrium phase separated system remains hyperuniform, even in the presence of attractive interactions. Fig.~\ref{fig: comparison}(e) shows the time evolution of $W^2(t)$ in both models for different system sizes starting from a flat interface at $t=0$. At equilibrium, as expected, we measure $W^2 \sim t^{2\chi/z}$ with $z=3$ ~\cite{onuki2002phase} for times before the saturation to a system size dependent plateau. For the hyperuniform system, however, the growth appears logarithmic as $\chi = 0$ and we cannot directly extract $z$ from $W^2(t)$. Instead, we can use the equilibration time $\tau_{\rm ss}(L) \sim L^z$ as demonstrated in Fig.~\ref{fig: comparison}(f), and the results are compatible with $z=3$.

 To verify the universality of our findings, we now consider a second class of hyperuniform models: random organization models. These models originated from the study of sheared colloids and represent perhaps the simplest framework for generating hyperuniform states. We study two variants in two dimensions~\cite{corte2008random,tjhung2015hyperuniform,PhysRevLett.131.047101, anand2025emergent}. The system consists of $N$ particles in a square box of side $L$ with periodic boundary conditions, using again a 50:50 binary mixture with two types of particles. To introduce attractive interactions, we generalize the original dynamic rule as follows. At each discrete time step $t$, particle pairs $(i,\,j)$ interact based on their pairwise distance $r_{ij}$ via two interaction ranges $\sigma_1$ and $\sigma_2>\sigma_1$. If $r_{ij}<\sigma_1$, particles undergo a repulsive displacement along the axis connecting the two centers, as in the original model. If $r_{ij} < \sigma_2$ and particles $(i,\,j)$ belong to the same species, they undergo an attractive displacement along the axis connecting their centers. During a collision, the amplitude of the resulting displacement is drawn randomly from uniform distributions in the intervals $[0, \epsilon_1]$ and $[0, \epsilon_2]$, respectively.

\begin{figure}[t]
\includegraphics[width=0.98\columnwidth,clip=true]{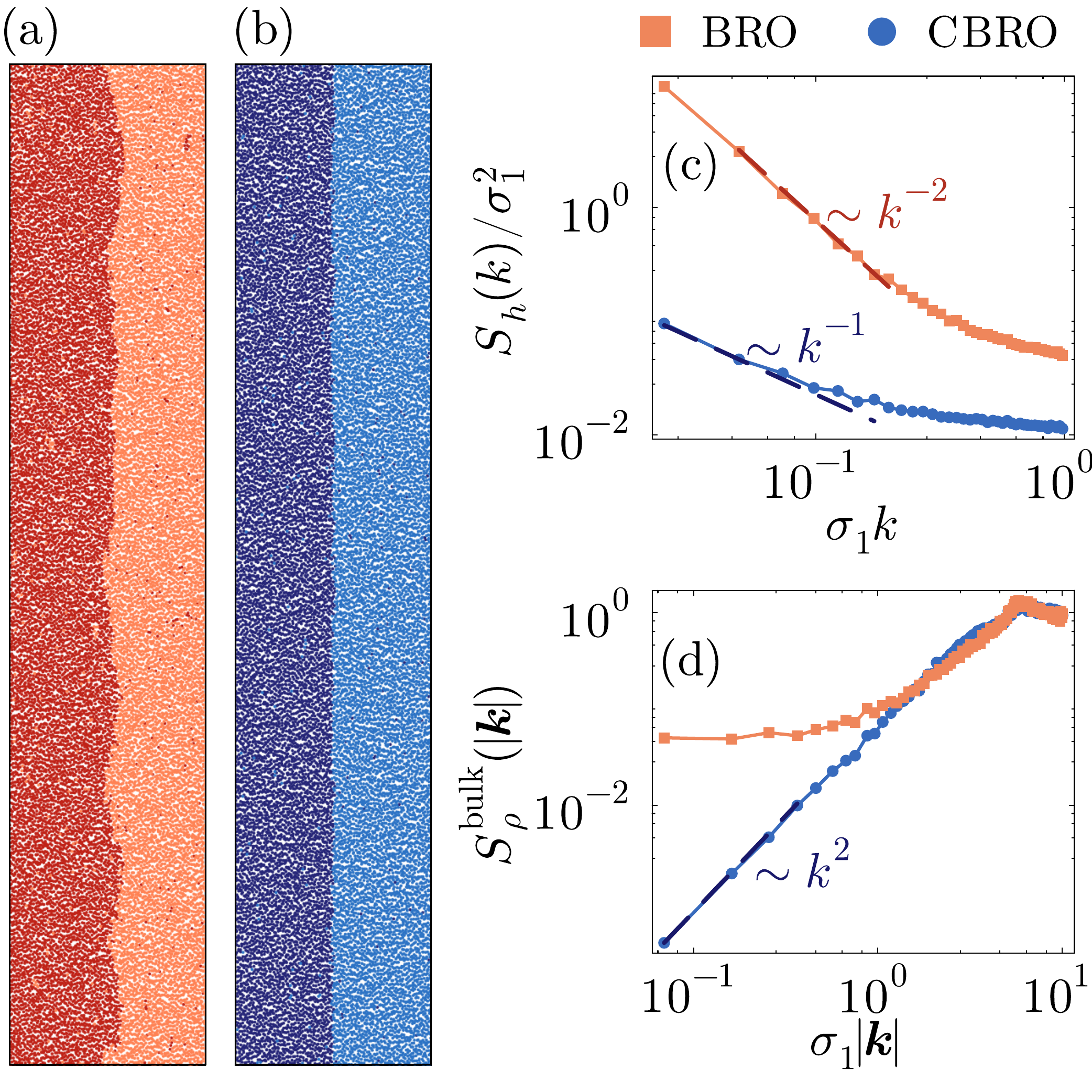} 
\caption{Random organization models. 
(a-b) Snapshots of a phase-separated mixtures evolving under BRO (a) and CBRO (b) dynamics. The parameters are: $\sigma_2/\sigma_1=1.4$, $\epsilon_1/\sigma_1=0.5$, $\epsilon_2/\sigma_1=0.1$, $N=5\times10^4$, $L=256\sigma_1$.
(c) Interface static height correlation functions.
(d) Radially averaged structure factors in the bulk. Results are averaged over 20 independent realizations.}   
\label{fig:rom_sketch} 
\end{figure}

In the conserved biased random organization model (CBRO), the displacement amplitudes are the same for the two interacting particles, so that the position of the center of mass is conserved~\cite{hexner2017noise, PhysRevLett.131.047101}. This conservation law is the source of the bulk hyperuniformity. To assess the role of hyperuniformity on the interfacial dynamics, we also study the biased  random organization model (BRO), where the displacement amplitudes of the two interacting particles are independent~\cite{wilken2021random}. This second model displays equilibrium-like density fluctuations and is not hyperuniform. 

The snapshots in Figs.~\ref{fig:rom_sketch}(a, b) show that the interface in the BRO model displays significantly more pronounced fluctuations compared to the CBRO model, where the interface appears nearly flat. The visual difference is quantitatively supported by the static height correlation $S_h(k)$ in Fig.~\ref{fig:rom_sketch}(c), where the BRO model shows a $|{\bm k}|^{-2}$ scaling, consistent with equilibrium-like fluctuations, whereas the CBRO model displays a milder divergence, again compatible with $|{\bm k}|^{-1}$. Finally, Fig.~\ref{fig:rom_sketch}(d) confirms the different nature of the density fluctuations in the bulk for the two models, even when attraction is present.  

\begin{figure}[t]
\includegraphics[width=0.98\columnwidth,clip=true]{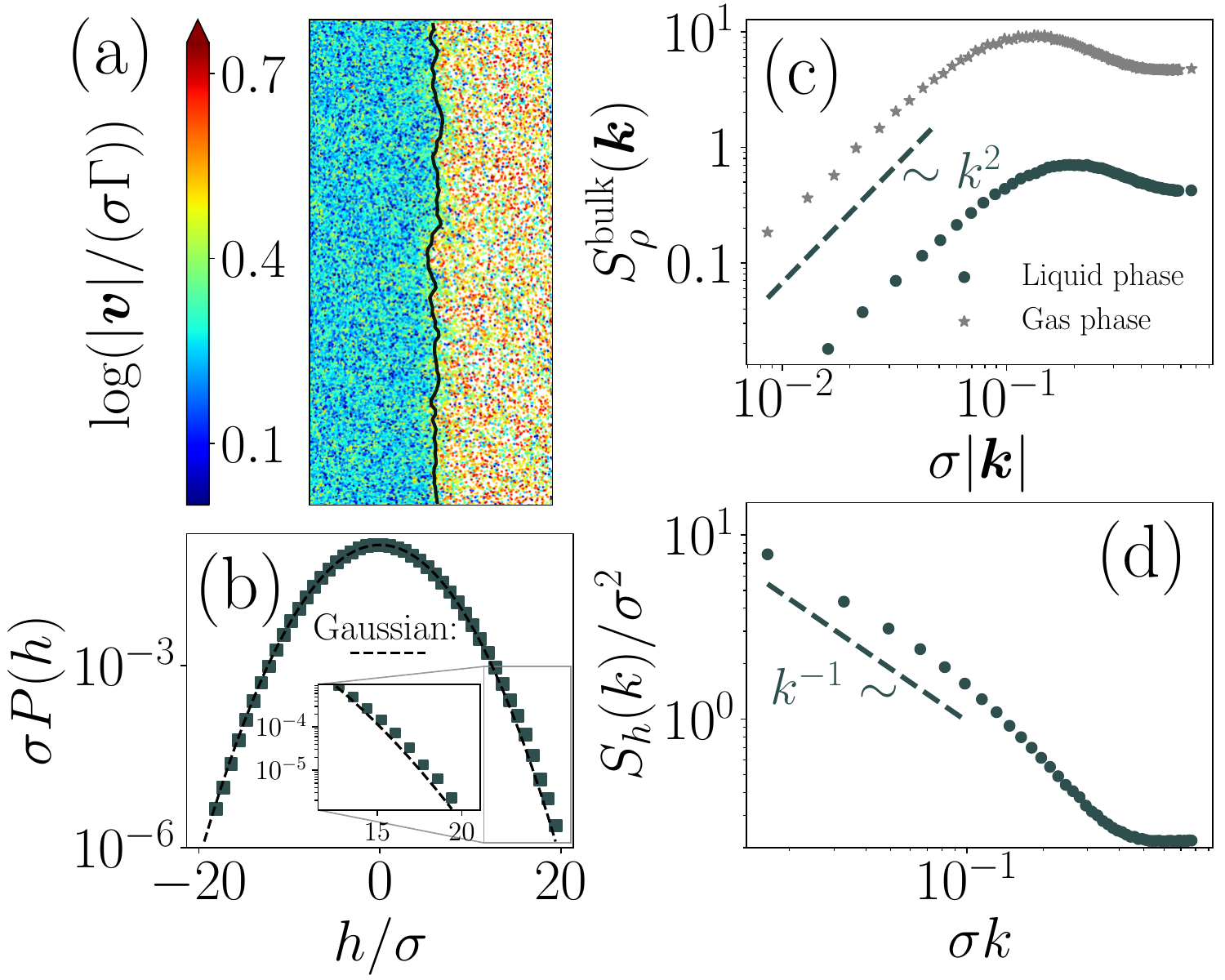} 
\centering
\caption{Mono-component system. (a) Snapshot with particles colored by velocity magnitude. (b) Probability distribution of height displacement $h$. The dashed line is a Gaussian fit. Inset: zoom on the distribution tail to reveal the asymmetry  of the distribution with measured skewness $\kappa_3\simeq0.045$ and excess kurtosis $\kappa_4\simeq0.031$ for this system size. (c) Structure factor of bulk coexisting liquid and gas phases. (d) Static height correlation. Parameters are $\delta E / \delta E_0 = 2 \times 10^3$, $\beta = 10$, $\tau_r\Gamma = 0.2$, $\delta E_0/m(\sigma\Gamma)^2=10/36$, $L_x=1.81L_y=383\sigma$ and $N\pi\sigma^2/(4L_xL_y)=0.31$ .}  
\label{fig: mono-component}
\end{figure}

Our third model starts from Eqs.~(\ref{eq: collision rule RanNi}, \ref{eq: damping}) in a mono-component system instead of a bidisperse one, but promotes $\Delta E$ to a function of the time elapsed since the last collision, effectively linking activity to the local density. Specifically, after a collision, particles undergo a recharging period before they can again transfer significant energy in subsequent collisions, so that less energy is injected in regions where collisions are more frequent. This may trigger a phase separation without attractive forces~\cite{risso2018effective}, in analogy with motility induced phase separation in active matter~\cite{cates2015motility}. Inspired by a model for vibrated granular materials, we choose~\cite{risso2018effective}:
\begin{equation}
\Delta E = 2\delta E_0 + \delta E\left(1 - e^{-\tau_i/\tau_r}\right)^\beta+\delta E\left(1 - e^{-\tau_j/\tau_r}\right)^\beta,   
\label{eq: mono-component collision rule}
\end{equation}
with $\tau_i$ the time since the last collision of particle $i$, 
$\tau_r$ a charging time, and $\beta$ controlling the form of the recharging function. The energy injected in a collision varies between $2\delta E_0$ and $2\delta E_0 + 2 \delta E$.

Fig.~\ref{fig: mono-component}(a) shows a typical snapshot of this model, with particles colored according to their velocities, demonstrating a phase separation between a dense phase with low kinetic energy, and a dilute phase with large velocities, consistent with underdamped non-equilibrium dynamics~\cite{maire2024non, hecht2024motility, mandal2019motility, caprini2024dynamical}. The difference in kinetic energy breaks the $h\to -h$ symmetry of the interface, as evidenced in Fig.~\ref{fig: mono-component}(b) showing that the height density distribution $P(h)$ is biased towards the dilute phase. Fig.~\ref{fig: mono-component}(c) confirms that both phases maintain hyperuniformity. Fig.~\ref{fig: mono-component}(d) demonstrates our main result: the non-equilibrium interface exhibits suppressed fluctuations compatible with $S_h(\bm k)\sim |\bm k|^{-1}$.

To explain the observed universal suppression of interfacial fluctuations, $S_h(\bm k) \sim |\bm k|^{-1}$, in the three proposed models, we consider a coarse-grained equation for the dynamics of a conserved scalar order parameter $\phi(\bm r,t)$, representing either the local concentration (liquid-liquid) or the local density (liquid-gas). We generalize the linear model from Ref.~\cite{hexner2017noise} to account for phase separation in center-of-mass conserving systems:
\begin{equation}
\begin{split}\partial_t\phi &= M \bm \nabla^2\dfrac{\delta F}{\delta \phi}+\sqrt{2 D}\bm\nabla^2\xi,\\
\frac{\delta F}{\delta \phi}&=-A \phi + A \phi^3-  \kappa\bm \nabla^2  \phi,
\end{split}
\label{eq: model B general modified} 
\end{equation}
where $F[\phi]$ is a free energy allowing for phase separation at $\phi=\pm1$~\cite{evans1979nature}, $\kappa>0$ adds cost to gradients and $A,  M, D > 0$ are positive parameters and $\xi$ is a random noise with zero mean and unit variance. Unlike equilibrium model B with noise term $\bm \nabla \cdot \boldsymbol\eta$ (conserving density but not center-of-mass), the noise $\bm \nabla^2 \xi$ preserves the position of the center-of-mass~\cite{de2024hyperuniformity}. Eq.~\eqref{eq: model B general modified} can be derived from the Navier–Stokes equations governing the density and velocity fields of the mono-component system described by Eqs.~(\ref{eq: collision rule RanNi},~\ref{eq: damping},~\ref{eq: mono-component collision rule}) (see Supplemental Material~\cite{supp}). For the other two models, involving binary mixtures, we assume Eq.~\eqref{eq: model B general modified} still applies when the system is fully phase-separated, thus neglecting interspecies diffusion. The binary mixture with damped velocity can also be described by an incompressible hydrodynamic equation instead of Eq.~\eqref{eq: model B general modified}, from which all the results presented below can likewise be derived, as shown in the Supplemental Material~\cite{supp}.  In the linear regime, Eq.~\eqref{eq: model B general modified} correctly predicts bulk hyperuniformity, $S_\rho(\bm k)\propto |\bm k|^2$~\cite{hexner2017noise}.

To derive an equation for the interface, we make the standard ansatz~\cite{bray2001interface}:
\begin{equation}
    \phi(\bm r, t)=f(r_\perp-  h(\bm r_\parallel, t)),
    \label{eq: ansatz}
\end{equation}
where $r_\perp$ and $\bm r_\parallel$ are the axis orthogonal and parallel to the interface, respectively. The function $f(x)$ varies sharply at the interface position, $x=0$. This ansatz projects the dynamics onto the interface $h$. By substituting Eq.~\eqref{eq: ansatz} into Eq.~\eqref{eq: model B general modified} and retaining only leading-order contributions, we obtain (see Supplementary Material~\cite{supp})  an equation for the interface dynamics:
\begin{equation}
    \partial_t h(\bm k, t) = -\dfrac{\gamma_0 M}{2}|\bm k|^3h(\bm k, t) + \sqrt{\dfrac{\gamma_0 D\bm k^2}{2\kappa}}\zeta(\bm k, t),
    \label{eq: interface dynamics}
\end{equation}
where $\zeta$ is a white noise with zero mean and unit variance, and $\gamma_0\equiv\kappa\int f'(r_\perp)^2dr_\perp$ the surface tension~\cite{onuki2002phase}. At equilibrium, the noise term in Eq.~(\ref{eq: interface dynamics}) would be proportional to $\sqrt{|\bm k|}\zeta$. This crucial difference in $k$ factors arises from the difference between the equilibrium divergence noise and the non-equilibrium rather Laplacian noise. Equation~\eqref{eq: interface dynamics} shows that non-equilibrium dynamics suppresses the height fluctuations leading to $\chi=(1-d)/2$ (instead of $\chi=(2-d)/2$ at equilibrium), so that 
\begin{equation} 
    S_h(\bm k) = \dfrac{ D}{2 M \kappa}|\bm k|^{-1}. 
    \label{eq: k^-1} 
\end{equation} 
This distinctive $|\bm k|^{-1}$ dependence, successfully observed across our three microscopic models, constitutes our main result. It shows that interface fluctuations between phase-separated hyperuniform phases are themselves strongly suppressed, thus leading to the concept of ``hyperuniform interfaces'' that are much flatter than their equilibrium counterparts~\footnote{The reduced fluctuations could likewise be characterized through the related notion of local surface fluctuation introduced in Ref.~\cite{torquato2016hyperuniformity}.}. For the dynamical exponent, we find $z=3$, in agreement with Fig.~\ref{fig: comparison}.

Finally, we discuss the influence of additional non-equilibrium effects. In Eq.~\eqref{eq: model B general modified}, non-equilibrium effects arise solely from the noise, while the deterministic part derives from a free energy. A non-equilibrium deterministic term in the order parameter equation would break the $h \to -h$ symmetry in the resulting interface equation~\cite{besse2023interface}. The asymmetric distribution of $h$ observed in Fig.~\ref{fig: mono-component} suggests the presence of such a term. The renormalization group analysis performed in the Supplementary Material~\cite{supp} shows that this contribution is at most RG-marginal in $d=1$, leading only to minor logarithmic corrections leaving the scaling behavior of $S_h$ predicted by the linear theory in Eq.~\eqref{eq: interface dynamics} unchanged. The marginality of the non-linearity can also account for the weakness of the observed non-Gaussianity of $P(h)$ in Fig.~\ref{fig: mono-component}(b).

In summary,  we have identified a new universality class for interfacial behavior driven by bulk hyperuniformity. Our simulations of three physically distinct models, inspired by vibrated granular media, sheared colloidal suspensions, and active systems with trapping mechanisms, consistently revealed a strong suppression of interfacial fluctuations, a result supported by our theoretical analysis. This shared
feature should lead to a finite interface width in $2d$. This stands in contrast to the logarithmic divergence with system size observed at equilibrium, and lowers the lower critical dimension of the roughening transition by one. This finding provides a new example showing how hyperuniformity dramatically impacts the nature of fluctuations in non-equilibrium systems. For instance, it reduces the upper critical dimensions  of $O(N)$ models \cite{gao2025liquidgascriticalityhyperuniformfluids}, and enables phase transitions in low-dimensional systems with $N>1$, bypassing the equilibrium lower critical dimension of 2~\cite{ikeda2023correlated}. Likewise, it enables a breakdown of the Mermin–Wagner theorem in two-dimensional crystals, allowing true translational long-range order~\cite{PhysRevLett.131.047101, ikeda2024harmonic,enhancing2024Maire, PhysRevE.104.064605, dey2024enhanced, kuroda2024long, keta2024long}. In all these cases, hyperuniformity efficiently suppresses the Goldstone modes arising from broken translational symmetry and decreases the lower critical dimension.

Because hyperuniform interfaces appear much smoother than equilibrium ones, this raises questions about the corresponding surface tension and possible physical consequences. Far from equilibrium, there exist multiple definitions of the surface tension. For example, using the capillary wave spectrum would directly lead to a diverging surface tension $\gamma \equiv \lim_{\bm k\to 0}T |\bm k|^{-2}/S_h(\bm k)\to\infty$, whereas this expression leads to a finite value at equilibrium. Instead, the non-equilibrium noise does not affect the surface tension $\gamma_0$ when using definitions based on Laplace pressure or the Buff–Kirkwood approach~\cite{braga2018pressure, ten1998computer}. This contrasts with findings in active matter~\cite{cates2024active, zhao2024active, zakine2020surface, langford2024mechanics}, where various novel effects are found from all definitions. 

In future work, it would be interesting to extend our analysis to a broader range of non-equilibrium situations, such as chiral particles~\cite{kuroda2023microscopic, kushwaha2024chirality, wang2025active, kuroda2025singulardensitycorrelationschiral}, cells~\cite{keta2024long, li2024fluidization} or other non-equilibrium systems~\cite{wiese2024hyperuniformity, boltz2024hyperuniformity, liu2025hyperuniform}. A second avenue is to broaden the search for peculiar interfacial effects beyond non‐equilibrium mechanisms. At equilibrium, long‐range forces generically produce hyperuniformity~\cite{leble2021two}, a peculiarity which also holds out-of-equilibrium~\cite{oppenheimer2022hyperuniformity, yashunsky2024topological}.  How such interactions reshape interface thermodynamics, the existence of a surface tension and fluctuations, remains almost entirely uncharted~\cite{mecke1999effective}. 

\begin{acknowledgments}
We thank G. Foffi and F. Smallenburg for their invaluable help and insightful discussions throughout this project. We also greatly benefited from exchanges with L. Chaix, M. Grant, Y. Kuroda, A. Maitra, C. Nardini, L. Sarfati, J. Tailleur, and J. Toner. L.B. acknowledges the support of the French Agence Nationale de la Recherche (ANR), under grants ANR-20-CE30-0031 (project THEMA) and ANR-24-CE30-0442 (project GLASSGO).
\end{acknowledgments}

\bibliography{bib}

\clearpage

\onecolumngrid

\appendix

\begin{center}
\Large{\textbf{Supplementary Material}}
\end{center}

\section{Derivation of the modified model B from hydrodynamics}\label{app:model B}

We outline how to derive a modified model B for the density field $\rho$ of a liquid-gas phase transition under non-equilibrium driving starting from an equation taking into account hydrodynamics. We begin with the following fluctuating compressible Navier-Stokes-Korteweg type equation with an Ekman damping $\overline\Gamma = \Gamma\rho$:
\begin{equation}
    \begin{split}
        \partial_t \rho + \bm\nabla \cdot ( \rho\bm v ) &= 0,\\
        \rho\left(\partial_t\bm v + (\bm v \cdot \bm \nabla)\bm v \right)&=-\overline\Gamma \bm v + \bm \nabla\cdot\left(\bm \sigma^{d} + \bm \sigma^\rho + \bm\sigma^r\right),\\
        \bm \sigma^d&= \eta(\bm\nabla \bm v + \bm  \nabla \bm v^\intercal) + \mu (\bm \nabla\cdot\bm v)\mathbb I,\\
         \langle\sigma_{ij}^r(\bm r, t)\sigma_{kl}^r(\bm r', t')\rangle &= 2 T \left[\eta\left(\delta_{ik}\delta_{jl} + \delta_{il}\delta_{jk}-\dfrac{2}{d}\delta_{ij}\delta_{kl}\right) +\mu\delta_{ij}\delta_{kl}\right]\delta(\bm r - \bm r')\delta(t - t'),\\
       \bm\nabla \cdot \bm \sigma^\rho &= -\rho\bm\nabla\dfrac{\delta F}{\delta \rho},\\
        F &= \int V(\rho) + \frac{\kappa}{2}(\bm\nabla\rho)^2 d\boldsymbol r.
    \end{split}
    \label{eq: hydro starting point}
\end{equation}
The first two equations show that $\rho$ is advected by $\bm v$, itself driven by stresses and damped by $\overline\Gamma$. $\bm \sigma^d$ is a dissipative stress arising from viscous forces with $\eta$ the shear viscosity and $\mu$ the bulk viscosity, $\bm \sigma^r$ is a random stress with variance postulated to satisfy the fluctuation-dissipation theorem with respect to $\bm \sigma^d$ (but not with $\overline\Gamma$) and $\bm\sigma^\rho$ is a reversible stress associated with gradients of the density field. In principle, the deterministic part of Eq.~\eqref{eq: hydro starting point} can be derived from the Boltzmann equation governing the evolution of dissipative hard disks~\cite{garzo2018enskog}. The modified Navier-Stokes-Korteweg Eq.~\eqref{eq: hydro starting point} then follows from a revised Chapman-Enskog procedure~\cite{dorfman2021contemporary}, in which the reversible stress is obtained by expanding the Boltzmann equation to second order in the Knudsen number while still neglecting Burnett corrections~\cite{giovangigli2020kinetic}. The temperature field is slaved to the density field and is therefore omitted in the Navier-Stokes equation.

On large length scales ($l^2\gg\eta/\overline\Gamma)$, terms involving derivatives of the velocity field will be negligible compared to $\overline\Gamma \bm v$ and the reversible stress. Therefore, Eq.~\eqref{eq: hydro starting point} simplifies to:

\begin{equation}
     \Gamma\partial_t \rho =\bm\nabla \cdot \left( \rho \bm\nabla\dfrac{\delta F}{\delta \rho}\right)-\bm\nabla\cdot(\bm\nabla\cdot\bm\sigma^r).
\end{equation}

Thanks to the autocorrelation of the random stress, the noise can be written as the Laplacian of a scalar random field. To recover an equation resembling model B, we simplify the expression $\rho(\bm r) \bm\nabla(\delta F/\delta \rho(\bm r))$ by replacing the density field with a constant average density $\rho_0$, yielding $\rho_0 \bm\nabla(\delta F/\delta \rho(\bm r))$. A similar approximation transforms the multiplicative noise--arising from the density dependence of viscosity and temperature--into additive noise by setting $\eta(\rho(\bm r)) \to \eta(\rho_0)$ and $\mu(\rho(\bm r)) \to \mu(\rho_0)$. Although uncontrolled, especially in inhomogeneous systems, this approximation greatly simplifies the analysis and is common in equilibrium settings. It underlies the derivation of model B and phase-field models from more microscopic theories, such as dynamical density functional theory or the Dean-Kawasaki equation~\cite{jacquin2015brownian, archer2019deriving, te2020classical, illien2024dean}. At equilibrium, despite modifying the dynamics, this approximation preserves the steady-state distribution $P_{\rm ss}[\rho] \propto e^{-F[\rho]/T}$, in compliance with the fluctuation-dissipation theorem and more surprisingly does not affect the universal coarsening~\cite{bray1994theory}.

This simplification leads to the modified model B introduced in the main text:
\begin{equation}
\begin{split}
\partial_t \rho &= M\bm\nabla^2\frac{\delta F}{\delta \rho} + \sqrt{2 D}\bm\nabla^2\xi,\\
F &= \int \left[ V(\rho) + \frac{\kappa}{2}(\bm\nabla\rho)^2 \right] d\boldsymbol r,
\end{split}
\label{eq: modified model B}
\end{equation}
where $\xi$ is a white Gaussian noise with unit variance, and $D$, $M > 0$ are constants. The system exhibits non-equilibrium dynamics due to the violation of the fluctuation-dissipation theorem; instead of the equilibrium $\bm\nabla \cdot \zeta$ noise, we find $\bm\nabla^2\xi$. This Laplacian form originates from the intrinsic stress-like character of the noise. Moreover, at mesoscopic scales, density fluctuations arise solely from collisions, which conserve both density and center-of-mass. Therefore, the form of Eq.~\eqref{eq: modified model B} is further justified by universality: it is the minimal stochastic extension of the Cahn–Hilliard equation that conserves both density and center-of-mass, consistent with the underlying physical constraints.

\section{Derivation of the dynamics of the interface for a liquid-gas phase separation from modified model B}\label{sec: modified model B interface dynamics}

To derive an interface equation, we start from the evolution of the order parameter:
\begin{equation}
    \begin{split}
    \partial_t \rho &= M\bm\nabla^2\frac{\delta F}{\delta \rho} + \sqrt{2D}\bm\nabla^2\xi,\\
        F &= \int V(\rho) + \frac{\kappa}{2}(\bm\nabla\rho)^2 d\boldsymbol r,\\
        V(\rho)&= A(\rho - \rho_l)^2(\rho - \rho_g)^2/4,
    \end{split}
    \label{eq: modified model B interface sec}
\end{equation}
%The local expansion of the free energy in power of gradient (as in usual $\phi^4$ theories) is justified by the short range nature of the interactions. However, at equilibrium, long-range weak potential such as $U(\bm r)\sim -|\bm r|^{-6}$ can hinder the convergence of this gradient expansion~\cite{evans1979nature}, leading to non-analyticity in the wavevector-dependent surface tension~\cite{mecke1999effective}. In what follows, we do not investigate the effect of such long-range forces on the non-equilibrium dynamics of our problem. 
and use the ansatz~\cite{bray2001interface}:
\begin{equation}
     \rho(\bm r, t)=f(r_\perp-  h(\bm r_\parallel, t)),
    \label{eq: Ansatz 2}
\end{equation}
where $h(\bm r_\parallel, t)$ denotes the interface position, $r_\perp$ is the coordinate normal to the interface, and $f$ describes the interfacial profile. At mean-field level, $f$ satisfies $\delta F / \delta \rho = \mu = 0$, yielding~\cite{onuki2002phase}:
\begin{equation}
f(x) \simeq f^{\text{MF}}(x) = \frac{\rho_l + \rho_g}{2} + \frac{\rho_l - \rho_g}{2} \tanh\left(\frac{x - x_0}{\delta}\right),
\label{eq: mean field f}
\end{equation}
with $\delta \propto \sqrt{\kappa / A}$ the interfacial width. In the sharp-interface limit, $f'$ is approximated by a Dirac delta of area $\rho_l - \rho_g$. The surface tension is found from the excess free energy stored in the interface \cite{onuki2002phase}: $\gamma_0 \equiv \kappa \int f'(u)^2  du$. %This will help us project the dynamics of the density onto the interface. Note that with this Ansatz, density fluctuations are excluded, meaning that sound wave contributions to the spectrum of the interfacial waves will not be captured. However, this is not a concern, as sound propagates much more quickly than capillary waves and are therefore negligible on large length scales~\cite{grant1983fluctuating}. 

Insertion of Eq.~\eqref{eq: Ansatz 2} into Eq.~\eqref{eq: modified model B interface sec} leads to:

\begin{equation}
    \begin{split}
        -\left[\partial_th(\bm r_\parallel, t)\right]f'(r_\perp-  h(\bm r_\parallel, t)) 
        =&M\bm\nabla^2\left[ \kappa h'' f'(r_\perp-  h(\bm r_\parallel, t))- \kappa(1+(h')^2)f''(r_\perp-  h(\bm r_\parallel, t))\right. \\
        &\left.+V'(f(r_\perp-  h(\bm r_\parallel, t)))\right] +\sqrt{2D}\bm\nabla^2\xi(\bm r_\parallel, r_\perp, t).
    \end{split}
    \label{eq: intermediate model b}
\end{equation}

We perform the useful change of variable $u = r_\perp - h(\bm r_\parallel, t)$ in Eq.~\eqref{eq: intermediate model b},  multiply formally by $f'(u)(\bm\nabla^2)^{-1}$ (justified by the vanishing of the deterministic part of the RHS at infinity) and integrate with respect to $u$:

\begin{equation}
    \begin{split}
        \int  f'(u)(-\bm\nabla^2)^{-1}\left(\left[\partial_th\right]f'(u)\right)du&=\int \left(\kappa M h'' f'(u)^2- \kappa M (1+(h')^2)f'(u)f''(u) +M V'(f(u))f'(u)\vphantom{\frac 2 1}\right.\\
        &~~\left.+\sqrt{2D}f'(u)\xi(\bm r_\parallel, r_\perp, t)\right)du\\
        &=\int \kappa M h'' f'(u)^2 -\sqrt{2D}f'(u)\xi(\bm r_\parallel, u, t)du + \mathcal{O}((\partial_{r_\perp}\xi) h)\\
        &\simeq M\gamma_0 \bm\nabla_\parallel^2 h + \sqrt{2D\gamma_0/\kappa}\zeta,
    \end{split}
    \label{eq: almost final model B}
\end{equation}
with $\zeta$ a white Gaussian noise with unit variance. We now compute $(-\bm\nabla^2)^{-1}$ by defining the Green function of minus the Laplacian: $G(r_\perp, \bm r_\parallel)$ for which $f(x)=(-\nabla^2)^{-1}g(x)=\int G(x-x') g(x') dx'$. Straightforward computations lead to $G(\bm k_\parallel, r_\perp-r'_\perp)=e^{-|\bm k_\parallel||r_\perp-r'_\perp|}/2|\bm k_\parallel|$. This simplifies the LHS of Eq.~\eqref{eq: almost final model B}:
\begin{equation}
    \begin{split}
        \int du f'(u)(-\nabla^2)^{-1}\left(\left[\partial_th(\bm k_\parallel)\right]f'(u)\right)&=\int du\int dv f'(u)G(\bm k_\parallel, u-v) f'(v)\left[\partial_th(\bm k_\parallel)\right]= \frac{\Delta \rho^2}{2|\bm k_\parallel|}\partial_th(\bm k_\parallel)+\mathcal{O}(\bm k_\parallel^0),
    \end{split}
    \label{eq: green function integration}
\end{equation}
with $\Delta\rho = \rho_l-\rho_g$ arising from the approximation $f'(u)\simeq\Delta\rho\delta(u)$.
Eqs~\eqref{eq: green function integration} and~\eqref{eq: almost final model B} lead to the interface equation in the monocomponent liquid-gas phase separated system:
\begin{equation}
    \partial_t h(\bm k, t) = -\dfrac{2M\gamma_0}{\Delta\rho^2}|\bm k|^3h(\bm k, t) + \sqrt{\dfrac{8D\gamma_0\bm k^2}{\kappa\Delta\rho^4}}\zeta(\bm k, t).
    \label{eq: liquidGasPhaseSeparation k-1 SI}
\end{equation}

\section{Derivation of the dynamics of the interface for a binary fluid mixture described via a damped model H} \label{app:model H}

We derive the interface dynamics for a conserved order parameter in the presence of hydrodynamic effects, using Model H, appropriate for incompressible $AB$ mixtures far from criticality.

 We introduce the concentration field $c\in [-1, 1]$:
\begin{equation}
    c(\bm r) = \dfrac{\rho_A(\bm r)-\rho_B(\bm r)}{\rho_A(\bm r) + \rho_B(\bm r)},
\end{equation}
with $\rho_A(\bm r)$ and $\rho_B(\bm r)$, the local densities of $A$ and $B$. Its evolution is postulated to follow a modified model H:
\begin{equation}
    \begin{split}
        \partial_t c + \bm v \cdot \bm \nabla c &= -\bm \nabla\cdot(-\mathcal M(c)\bm \nabla \mu + \bm{\mathcal J}^r),\\
        \rho\left(\partial_t\bm v + (\bm v \cdot \bm \nabla)\bm v \right)&=-\Gamma \rho\bm v + \eta \bm\nabla^2\bm v-\boldsymbol\nabla p - c \bm \nabla \mu +\bm \nabla \cdot \bm\sigma^r, \quad\bm \nabla \cdot \bm v = 0\\
        \langle\sigma_{ij}^r(\bm r, t)\sigma_{kl}^r(\bm r', t')\rangle &= 2 \eta T [\delta_{ik}\delta_{jl} + \delta_{il}\delta_{jk}]\delta(\bm r - \bm r')\delta(t - t'),\\
        \mu \equiv \frac{\delta F}{\delta c},\quad F &= \int V(c) + \frac{\kappa}{2}(\bm\nabla c)^2 d\boldsymbol r,\quad     V(c)=A(c-1)^2(c+1)^2/4.
    \end{split}
    \label{eq: model H appendix}
\end{equation}
The first equation describes concentration evolution with advection by velocity $\bm v$, and diffusion with $\mathcal M$ a mobility and $\bm{\mathcal J}^r$ a random current. The second equation is the incompressible fluctuating Navier-Stokes equation with global damping $\Gamma$, viscous damping $\eta\bm\nabla^2 \bm v$ and coupling to the concentration field $c$ via the chemical potential $\mu$ defined from a free energy functional $F$.

%Since the system is out of equilibrium, the underlying field theory governing its dynamics may contain terms that do not derive from a free energy, as seen in active Model H~\cite{tiribocchi2015active}. Here, we do not explicitly include such non-conservative terms. Instead, we will introduce potential nonlinearities that could emerge from these terms in the final interface equation and assess their relevance using the renormalization group approach in an other section.

Interspecie diffusion is confined to the interface region where $|c| \ll 1$, which in a low-temperature, phase-separated system is negligible: $-\mathcal M(c)\bm \nabla \mu + \bm{\mathcal J}^r\simeq0$. %This is justified since interface motion is primarily governed by non-local effects from the bulk phases~\cite{grant1983fluctuating}. At equilibrium, the mobility $M(c)$ and the autocorrelation of $\bm {\mathcal J}^r$ are related to the chemical potential via the Onsager relation, with $\mathcal M(c) \propto (\partial_c \mu)^{-1}$~\cite{de2006hydrodynamic}. However, our $c^4$ theory yields $M(c = \pm 1) \neq 0$, which is inconsistent and permits unphysical values $|c| > 1$. This issue would be resolved using a Flory–Huggins free energy, for which $M(c = \pm 1) = 0$\cite{de2006hydrodynamic}.
Moreover, assuming small-amplitude interfacial fluctuations~\cite{grant1983fluctuating}, we neglect both the convective term and time derivative in the velocity equation, since $\bm v$ relaxes quickly due to damping. This reduces Eqs.~\eqref{eq: model H appendix} to a linear Stokes equation:
\begin{equation}
    \begin{split}
    \bm \nabla \cdot \bm v &= 0,\\
        -\overline\Gamma \bm v + \eta \nabla^2\bm v-\boldsymbol\nabla p &= \bm f(c) \equiv  c \bm \nabla \mu +\bm \nabla \cdot \bm\sigma^r,
    \end{split}
    \label{eq: stokes}
\end{equation}
with $\overline\Gamma = \Gamma\rho$ and $\bm f(c)$ driving the velocity field. The Stokes equation~\eqref{eq: stokes} is solved using a modified Oseen tensor $\mathbb T (\bm r)$, defined in Fourier space as $\mathbb T(\bm k)$ which verifies:
\begin{equation}
    \bm v(\bm k)=\dfrac{1}{\overline\Gamma + \eta \bm k^2}\left(\mathbb{I}-\frac{\bm k \otimes\bm k}{\bm k^2} \right)\bm f(\bm k)\equiv \mathbb T(\bm k)\bm f(\bm k).
    \label{eq: oseentensordefinitionofvelocity}
\end{equation}

We now derive the final equation governing the concentration field in the phase-separated system. Starting from Eq.~\eqref{eq: model H appendix} and substituting the velocity field via the Oseen tensor from Eq.~\eqref{eq: oseentensordefinitionofvelocity} (with repeated indices implying summation), we obtain:
\begin{equation}
    \begin{split}
        \partial_t  c &=-\int \left[\partial_i  c(\bm r) \mathbb T_{ij}(\bm  r'-\bm  r) \partial_i c(\bm r') \right]  \mu(\bm r') d\bm r'+\zeta(\bm r),\\
        %\zeta(\bm r) &= \int \partial_i  c(\bm r) \mathbb T_{ij}(\bm r-\bm r')\partial_k\sigma_{kj}^r(\bm r')d\bm r',\\
        \langle\zeta(\bm r, t) \zeta(\bm r', t')\rangle&=2 T \partial_i c(\bm r) \tilde {\mathbb T}_{ij}(\bm r - \bm r')\partial_j c(\bm r')\delta(t-t'),
    \end{split}
    \label{eq: model H final}
\end{equation}
with 
\begin{equation}
    \tilde {\mathbb T}(\bm k)\equiv \eta\bm k^2\text{Tr}\left[\mathbb T(\bm k)\mathbb T^\intercal(\bm k)\right]=\dfrac{\eta\bm k^2}{(\overline\Gamma + \eta \bm k^2)^2}\left(\mathbb{I}-\dfrac{\bm k\otimes \bm k}{\bm k^2}\right).
\end{equation}

In contrast to equilibrium systems ($\overline\Gamma = 0$)~\cite{bray2001interface} and other non-equilibrium cases~\cite{caballero2024interface}, $\tilde{\mathbb T} \neq \mathbb T$ due to the violation of the fluctuation-dissipation theorem: the damping $\overline\Gamma$ in our non-equilibrium setting lacks a corresponding noise.

In order to obtain an evolution equation for the interface, we use again the ansatz:
\begin{equation}
    c(\bm r, t)=f(r_\perp-  h(\bm r_\parallel, t)).
    \label{eq: Ansatz}
\end{equation}

Replacing Eq.~\eqref{eq: Ansatz} in Eq.~\eqref{eq: model H final}, we obtain:
\begin{equation}
    \begin{split}
 -(\partial_t h(\bm r_\parallel))f'(r_\perp-  h(\bm r_\parallel, t))
 = &-\int \left[\partial_i f(r_\perp-  h(\bm r_\parallel, t)) \mathbb T_{ij}(\bm  r'_\parallel-\bm  r_\parallel, r_\perp - r'_\perp) \partial_if(r_\perp'-  h(\bm r_\parallel', t)) \right]\\
 &\times\left[\kappa h'' f'(r_\perp'-  h(\bm r_\parallel', t))-  \kappa(1+(h')^2)f''(r_\perp'-  h(\bm r_\parallel', t))+V'(f(r_\perp'-  h(\bm r_\parallel', t))) \right] d\bm r'\\&+\zeta(\bm r_\parallel, r_\perp).
\end{split}
\label{eq: interface dynamic begninng}
\end{equation}

We perform again the useful change of variable $u = r_\perp - h(\bm r_\parallel, t)$, and use the standard approximations: 
\begin{equation}
    \begin{split}
    \partial_if(u)&\simeq f'(u)\delta_{i\perp},\\
     \mathbb T_{\perp \perp}(\bm  r'_\parallel-\bm  r_\parallel, u -u' - h(r_\parallel) - h(r_\parallel'))&\simeq \mathbb T_{\perp \perp}(\bm  r'_\parallel-\bm  r_\parallel, u -u'),\\
     \zeta(\bm r_\parallel, u+h(r_\perp))&\simeq\zeta(\bm r_\parallel, u).
    \end{split}
\end{equation}
We multiply both sides of Eq.~\eqref{eq: interface dynamic begninng} by $f'(u)$ and integrate afterward along $u$:
\begin{equation}
    \begin{split}
         \gamma_0\partial_t h(\bm r_\parallel)/\kappa
         \simeq& -\int \left[ \left(\int f'(u, t)^2 \mathbb T_{\perp \perp}(\bm  r'_\parallel-\bm  r_\parallel, u -u') du\right) f'(u', t)) \right]\\
         &\times\left[-\kappa h'' f'(u', t)+ \kappa (1+(h')^2)f''(u', t)-V'(f(u', t)) \right] d\bm r'+\int f'(u)\zeta(\bm r_\parallel, u)du\\
         \simeq&~\gamma_0^2\int \mathbb T_{\perp\perp}(\bm  r'_\parallel-\bm  r_\parallel, r_\perp=0)\nabla^2_\parallel h(\bm  r'_\parallel)d\bm r_\parallel'/\kappa+ \gamma_0\xi(\bm r_\parallel)/\kappa.
    \end{split}
    \label{eq: almost final model h interface}
\end{equation}
The white Gaussian noise $\xi$ has 0 average and the following correlation at lowest order:
\begin{equation}
    \langle\xi(\bm k_\parallel, t) \xi(\bm k_\parallel', t')\rangle=2 T\tilde {\mathbb T}_{\perp\perp}(\bm k_\parallel, r_\perp=0)(2\pi)^{d}\delta(\bm k_\parallel+\bm k_\parallel')\delta(t-t'),
\end{equation}
with $d$ the dimension of the interface (i.e., the dimension of the system minus one).

We now evaluate the Oseen tensors. We recall that
\begin{equation}
\begin{split}
    \mathbb T_{\perp\perp}(\bm k_\parallel, k_\perp)&=\dfrac{1}{\overline\Gamma + \eta (\bm k_\parallel^2 + k_\perp^2)}\left(1-\dfrac{k_\perp ^2}{\bm k_\parallel^2 + k_\perp^2}\right),\\
    \tilde {\mathbb T}_{\perp\perp}(\bm k_\parallel, k_\perp)&=\dfrac{\eta(\bm k_\parallel^2 + k_\perp^2)}{(\overline\Gamma + \eta (\bm k_\parallel^2 + k_\perp^2))^2}\left(1-\dfrac{k_\perp^2}{\bm k_\parallel^2 + k_\perp^2}\right).
\end{split}
\end{equation}

A partial Fourier transform leads to:
\begin{equation}
\begin{split}
    {\mathbb T}_{\perp\perp}(\bm k_\parallel, r_\perp)&=\dfrac{|\bm k_\parallel|}{2\overline\Gamma}\left(e^{-|\bm k_\parallel| |r_\perp|}-|\bm k_\parallel|\dfrac{1}{\sqrt{\overline\Gamma/\eta + \bm k_\parallel^2}}e^{-\sqrt{\overline\Gamma/\eta +\bm k_\parallel^2}|r_\perp|}\right),\\
    \tilde {\mathbb T}_{\perp\perp}(\bm k_\parallel, r_\perp)&=\dfrac{\bm k_\parallel^2e^{-\sqrt{\overline\Gamma/\eta}|r_\perp|}}{4(\overline\Gamma + \bm k_\parallel^2\eta)}\left(|r_\perp|+\dfrac{1}{\sqrt{\overline\Gamma/\eta+\bm k_\parallel^2}}\right).
\end{split}
\end{equation}

Therefore, for the long-wavelength dynamics, we obtain:
\begin{equation}
\begin{split}
    {\mathbb T}_{\perp\perp}(\bm k_\parallel, r_\perp=0)&=\dfrac{|\bm k_\parallel|}{2\overline\Gamma}\left(1-|\bm k_\parallel|\dfrac{1}{\sqrt{\overline\Gamma/\eta + \bm k_\parallel^2}}\right)=\dfrac{1}{2\overline\Gamma} |\bm k_\parallel|+\mathcal{O}(\bm k_\parallel^2),\\
    \tilde {\mathbb T}_{\perp\perp}(\bm k_\parallel, r_\perp=0)&=\dfrac{\bm k_\parallel^2}{4(\overline\Gamma + \bm k_\parallel^2\eta)\sqrt{\overline\Gamma/\eta+\bm k_\parallel^2}}=\dfrac{\sqrt{\eta/\overline\Gamma}}{4\overline\Gamma} \bm k_\parallel^2 + \mathcal{O}(\bm k_\parallel^4).
\end{split}
\label{eq: scaling}
\end{equation}

Using Eqs~\eqref{eq: scaling} and~\eqref{eq: almost final model h interface}, we finally obtain our final equation for the dynamics of the interface on large wavelength:
\begin{equation}
    \partial_t h(\bm k, t)=-\dfrac{\gamma_0}{2\overline\Gamma}|\bm k|^3h(\bm k, t)+\sqrt{\dfrac{T\sqrt{\eta/\overline\Gamma}}{2\overline\Gamma}\bm k^2}\zeta~~~~~~~|\bm k|\ll \sqrt{\overline\Gamma/\eta},
    \label{eq: liquidLiquidPhaseSeparation k-1 SI}
\end{equation}
with $\zeta$ a white Gaussian noise with unit variance. Compared to an equilibrium dynamics when the bulk is diffusive, the noise is proportional to $|\bm k|$ instead of $\sqrt{|\bm k|}$. 

We note that for intermediate $\bm k$, an effective equilibrium is recovered as:
\begin{equation}
     \tilde {\mathbb T}_{\perp\perp}(\bm k_\parallel, r_\perp=0)={\mathbb T}_{\perp\perp}(\bm k_\parallel, r_\perp=0) + \mathcal{O}(|\bm k_\parallel|^{-3})=\dfrac{1}{4\eta |\bm k_\parallel|}+\mathcal{O}(|\bm k_\parallel|^{-3}).
\end{equation}
Therefore, the equation for the dynamics of $h$ for large $\bm k$ reads:
\begin{equation}
    \partial_t h(\bm k, t)=-\dfrac{\gamma_0}{4\eta}|\bm k|h(\bm k, t)+\sqrt{\dfrac{T}{2\eta |\bm k|}}\zeta(\bm k, t)~~~~~~~|\bm k|\gg \sqrt{\overline\Gamma/\eta}.
\end{equation}
This results correspond to the one obtained at equilibrium when the dynamics of the interface is completely controlled by hydrodynamics effects \cite{bray2001interface}. This is not a surprise as the dynamics of the large $\bm k$ modes if effectively unscreened, and the fluctuation-dissipation theorem for these modes of the velocity field is approximately recovered since $\overline\Gamma$ is negligible.

\section{Renormalization group analysis of the non-linear interface equation\label{sec: RG}}

A new universality class for non-equilibrium interfaces called $|\bm q|$KPZ was recently found due to the possible existence of the non-linearity $-\lambda |\bm k|\int_{|\bm q| <\Lambda}\bm q\cdot(\bm q-\bm k)h(\bm q, t)h(\bm q - \bm k, t)d\bm q/(2\pi)^d$ in the equation of motion of $h$, with $\lambda$ a coupling constant and $\Lambda$ an ultraviolet cut-off. It is not derivable from a free energy and breaks the equilibrium $h\to-h$ symmetry. Introducing non-free energy terms or multiplicative noise in the order parameter equation is expected to generate this term in our interface equation~\cite{besse2023interface, caballero2024interface}. 

In $1d$, this non-linearity is marginal for our system (relevant for Ref.~\cite{besse2023interface}), suggesting logarithmic corrections to the linear theory's scaling. While analogous systems can exhibit strong-coupling behavior that alters this conclusion~\cite{kardar1986dynamic, caballero2018strong}, our one-loop calculation reveals no such strong-coupling regime for our hyperuniform interface. We do, however, confirm the presence of logarithmic corrections as expected.

We start from our interface equation and add the non-linearity:
\begin{equation}
    \partial_t h(\boldsymbol k, t)=-\gamma |\boldsymbol k|^3h(\boldsymbol k, t) - \lambda |\bm k|\int_{|\bm q|<\Lambda}\bm q\cdot  (\bm q - \bm k)h(\bm q, t)h(\bm q-\bm k, t)\dfrac{d\bm q}{(2\pi)^d}+\tilde\eta(\boldsymbol k, t),
    \label{eq: starting point non linear study}
\end{equation}
with:
\begin{equation}
    \langle\tilde\eta(\bm k, t)\rangle=0~~~~~~~~~~~~~\langle \tilde\eta(\bm k, t)\tilde\eta(\bm k', t')\rangle=(2\pi)^d2D\delta(\bm k + \bm k')\delta(t-t').
\end{equation}
We rearrange Eq.~\eqref{eq: starting point non linear study} in a more appropriate way to perform a perturbative expansion~\cite{toner2024physics}:
\begin{equation}
    \begin{split}
        h(\boldsymbol k, w)&= G_0(\bm k, w)\tilde\eta(\boldsymbol k, t)+  G_0(\bm k, w)\int_{|\bm q| < \Lambda}\int_{-\infty}^{\infty} g_0(\bm k, \bm q,\bm q - \bm k)h(\bm q, \Omega)h(\bm q-\bm k, \Omega - w)\dfrac{d\bm q}{(2\pi)^d}\dfrac{d\Omega}{2\pi},\\
    \end{split}
    \label{eq: beginning pertubation serie}
\end{equation}
with:
\begin{equation}
    \begin{split}
        G_0(\bm k, w)=\dfrac{1}{-iw+\gamma|\bm k|^3}, \quad
        g_0(\bm k, \bm q, \bm p)&=-\lambda |\bm k|\bm q\cdot \bm p,\quad
    C_0(\bm k, w)%=\dfrac{G_0(\bm k, w)\langle \tilde \eta(\bm k, w)\tilde\eta(\bm k', w')\rangle G_0(\bm k', w')}{(2\pi)^{d+1}\delta(w+w')\delta(\bm k+\bm k')}
    =\dfrac{2D|\bm k|^2}{w^2+\gamma^2|\bm k|^6},
    \end{split}
    \label{eq: bare parameter}
\end{equation}
where $G_0$, $g_0$ and $C_0$ are the bare propagator, vertex and correlator, respectively.

Successive insertion of the left-hand side of Eq.~\eqref{eq: beginning pertubation serie} in its right-hand side, leads to a perturbative solution of the equation. Using standard method we can organize the perturbation expansion in Feynman diagrams where: 
$G_0(\bm k, w)=\vcenter{\hbox{\raisebox{2.5ex}{\includegraphics{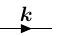}}}}$
, $g_0(\bm k, \bm q, \bm k-\bm q) =\vcenter{\hbox{\raisebox{-1ex}{\includegraphics{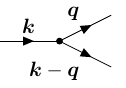}}}}$ 
and $C_0(\bm k, w)=\vcenter{\hbox{\raisebox{3ex}{\includegraphics{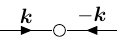}}}}$

For clarity, frequencies are omitted from the diagrams. The complete diagrams include loops where internal momenta are integrated within the high-energy shell  $\Lambda(1-\delta l)\leq |\bm q| \leq \Lambda$. This restriction avoids ultraviolet and infrared divergences. These integrations will partially re-sum the bare parameters, leading to the Wilson renormalization group. We set the external frequency to 0 ($w=0$) when calculating renormalized quantities, as higher time derivatives are irrelevant.

The propagator gets the following contribution:
\begin{equation}
    G_I(\bm k, 0) = G_0(\bm k, 0) + 4 \times \vcenter{\hbox{\raisebox{2.2ex}{\includegraphics{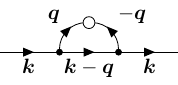}}}}.
\end{equation}

Using the rules introduced above, the integral over the internal frequency $\Omega$ is performed on $\mathbb R$ as it is convergent and the integral over momenta on $\Lambda(1-\delta l)\leq |\bm q| \leq \Lambda\equiv \text{shell}$, to eliminate the fine scales:
\begin{equation}
    \begin{split}
        G_I(\bm k, 0) &= G_0(\bm k, 0)+4G_0(\bm k, 0)^2\int_{-\infty}^\infty\int_\text{shell} g_0(\bm k, \bm q, \bm k - \bm q)G_0(\bm k-\bm q, -\Omega)C_0(\bm q, \Omega)g_0(\bm k-\bm q, -\bm q, \bm k)\dfrac{d\bm q d\Omega}{(2\pi)^{d+1}}\\
        %&=G_0(\bm  k, 0)\left(1-\dfrac{4\lambda^2 D}{\gamma^3}\int_\text{shell} \dfrac{|\bm k||\bm k-\bm q| (\bm q\cdot (\bm k-\bm q))\left(\bm q\cdot\bm k\right)}{|\bm k|^3|\bm q|(|\bm q|^3 + |\bm q - \bm k|^3)}   \dfrac{d \bm q}{(2\pi)^d} \right)\\
        &= G_0(\bm  k, 0)\left(1-\dfrac{\lambda^2 D}{\gamma^3}\int_\text{shell} \dfrac{(\bm q\cdot\bm k)^2}{|\bm k|^2|\bm q|^3}   \dfrac{d \bm q}{(2\pi)^d} +\mathcal{O}(|\boldsymbol k|)\right).
    \end{split}
\end{equation}
As $G_0^{-1}(\bm k, 0)=\gamma |\bm k|^3$, $\gamma$ gets renormalized as follows:
\begin{equation}
    \gamma_I \simeq \gamma\left(1+\dfrac{\lambda^2D}{\gamma^3}\dfrac{K_d}{d}\Lambda^{d-1}\delta l\right),
\end{equation}
with $K_d=S_d/(2\pi)^d$ and $S_d$ the area of the unit sphere in $d$ dimensions. Where $\gamma_I$ is the new ``dressed" surface tension after the integration of the high momenta modes.

$D$ is renormalized through the renormalization of the correlator $C$ at $w=0$ at one loop: 
\begin{equation}
    C_I(\bm k, 0) = C_0(\bm k, 0) + 2 \times \vcenter{\hbox{\raisebox{0ex}{\includegraphics{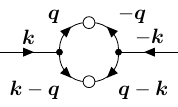}}}}+4\times \vcenter{\hbox{\raisebox{2.2ex}{\includegraphics{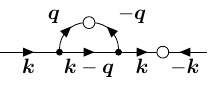}}}}+4\times \vcenter{\hbox{\raisebox{2.2ex}{\includegraphics{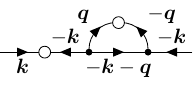}}}}.
    \label{eq: renormalization correlator}
\end{equation}
The last two diagrams are already included in the renormalization of $G$. The first diagram is more interesting and corresponds to a renormalization of $D$:
\begin{equation}
    \begin{split}
        C_I(\bm k, 0)&=C_0(\bm k, 0)+2\left|G_0(\bm k, 0)\right|^2\int_{-\infty}^\infty\int_\text{shell} g_0(\bm k, \bm q, \bm k - \bm q)C_0(\bm k-\bm q, -\Omega)C_0(\bm q, \Omega)g_0(-\bm k, -\bm q, \bm q - \bm k)\dfrac{d\bm q d\Omega}{(2\pi)^{d+1}}+\mathcal G_G\\
        %&=\left|G_0(\bm k, 0)\right|^2\left(2D|\bm k|^2 + \frac{4\lambda^2D^2}{\gamma^3}|\boldsymbol k|^2\int_\text{shell} \dfrac{(\bm q\cdot(\bm q - \bm k))^2}{|\bm k -\bm q||\bm q|\left(|\bm k -\bm q|^3 + |\bm q|^3\right)}\frac{ d \bm q}{(2\pi)^d}\right)+\mathcal G_G\\
        &=\left|G_0(\bm k, 0)\right|^2\left(2D|\bm k|^2 + 2D|\bm k|^2\frac{\lambda^2D}{\gamma^3}\int_\text{shell}|\bm q|^{-1}\frac{ d \bm q}{(2\pi)^d}+\mathcal{O}(|\boldsymbol k|^4)\right)+\mathcal G_G,
    \end{split}
\end{equation}
where $\mathcal G_G$ correspond to the last two diagrams of Eq.~\eqref{eq: renormalization correlator}. Therefore, $D$ gets the following contribution:
\begin{equation}
    D_{I}\simeq D\left(1+\dfrac{\lambda^2D}{\gamma^3}K_d\Lambda^{d-1}\delta l\right).
\end{equation}

The graphs contributing to the renormalization of the vertex are:
\begin{equation}
    \begin{split}
    \mathcal{G}_{g_1}(\bm k, \bm p) = 4 \times\vcenter{\hbox{\raisebox{0ex}{\includegraphics{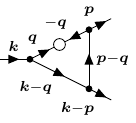}}}},\quad
    \mathcal{G}_{g_2}(\bm k, \bm p) = 4 \times \vcenter{\hbox{\raisebox{0ex}{\includegraphics{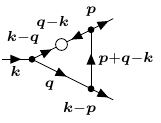}}}},\quad
    \mathcal{G}_{g_3}(\bm k, \bm p) = 4 \times \vcenter{\hbox{\raisebox{0ex}{\includegraphics{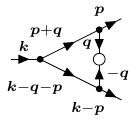}}}}.
    \end{split}
\end{equation}
Direct computations show that at the lowest order, the non-linearity $\lambda$ does not get renormalized at one loop because $\mathcal{G}_{g_1}(\bm k, \bm p)+\mathcal{G}_{g_2}(\bm k, \bm p)+\mathcal{G}_{g_3}(\bm k, \bm p)=\mathcal{O}(|\bm k|^3, \bm p^2)$ . This is usual in conserved KPZ like equations~\cite{besse2023interface, sun1989dynamics, caballero2024interface}. Note that the non-linearity is not protected from renormalization by a Galilean invariance as is the case for the KPZ equation and therefore, probably does get renormalized at two loops~\cite{janssen1997critical}.

In order to complete the renormalization procedure, we rescale the parameters and the field to restore the cut-off to $\Lambda$. This gives us the RG flow for $D$, $\gamma$ and $\lambda$:
\begin{equation}
    \begin{split}
        \dfrac{d \lambda}{dl}&=(z-3+\chi)\lambda,\\
        \dfrac{d \gamma}{dl}&= \left(z-3+\dfrac{\lambda^2 D}{d\gamma^3}K_d\Lambda^{d-1}\right)\gamma, \\
        \dfrac{d D}{dl}&= \left(z-2\chi-d-2+\dfrac{\lambda^2 D}{\gamma^3}K_d\Lambda^{d-1}\right)D.
    \end{split}
\end{equation}

It is however convenient to define the adimensional quantity $g=(\lambda^2D/\gamma^3)\Lambda^{d-1}K_d$ which quantifies the importance of the non-linearity in the system. The RG flow of $g$ at first loop is:
\begin{equation}
    \dfrac{d g}{dl}=\left(1-d-\dfrac{3-d}{d}g\right)g.
    \label{eq: RG flow}
\end{equation}
As dimensional analysis predicts, the Gaussian fixed point remains stable in $d \leq 2$, with no evidence of strong coupling at one-loop. In $d=1$ however, the non-linearity is marginal and therefore, logarithmic corrections can be expected. To show this, we follow Ref.~\cite{toner2023roughening}.

In $d=1$, the RG flow Eq.~\eqref{eq: RG flow} decays algebraically instead of exponentially in higher dimensions:
\begin{equation}
    \dfrac{d g}{dl}=-2g^2\Rightarrow g(l)= \dfrac{1}{2l+1/g_0}\qquad\text{ in } d=1,
    \label{eq: renormalized g}
\end{equation}
with $g_0$ the bare value of $g$. From the RG flow of $\gamma$ and $D$ in $1d$ and Eq.~\eqref{eq: renormalized g}, we obtain the values of $\gamma(l)$ and $D(l)$ at different RG ``time":
\begin{equation}
    \begin{split}
    \dfrac{d \gamma}{dl}\simeq(z-3+g(l))\gamma\Rightarrow \gamma(l) = \gamma_0\sqrt{1+2g_0l},\\
    \dfrac{d D}{dl}\simeq(z-2\chi - 2 - d+g(l))\gamma\Rightarrow \gamma(l) = D_0\sqrt{1+2g_0l},
    \end{split}
\end{equation}
where we choose $z=3$ and $\chi=(1-d)/2$. Therefore, while the non-linearity decays to zero under the RG flow, its marginality leads to a slow (algebraic rather than exponential) decay, which causes $\gamma$ and $D$ to diverge with $l$. In contrast, an exponential decay of the non-linear term would result in finite values of $\gamma(l)$ and $D(l)$ as $l \to \infty$. 

From the scaling of $h(\bm k, w)$ and the relation Eq.~\eqref{eq: bare parameter}, we can write down a relation concerning the exact correlation function $C$ at different step of the RG: 
\begin{equation}
    C(e^l \bm k, e^{zl}w, \gamma(l), D(l), \lambda(l))=e^{-(2\chi + z + d)l}C(\bm k, w, \gamma_0, D_0, \lambda_0).
    \label{eq: trajectory matching exact relation}
\end{equation}

This relation holds exactly when evaluated at a fixed point (otherwise, the exponents acquire a scale dependence on $l$). The parameter $l$ can be chosen arbitrarily, and a particularly convenient choice is $|\bm k| e^l=\Lambda$. With this choice, the left-hand side of Eq.~\eqref{eq: trajectory matching exact relation} is evaluated at the cutoff wavenumber $\Lambda$ and $l$ is large enough for the reduced nonlinearity to have become negligible. Consequently, up to small finite corrections arising from standard perturbation theory, the correlation function on the left-hand side can be accurately approximated by its counterpart from the linear theory. We obtain:
\begin{equation}
    \begin{split}
    C(\bm k, w, \gamma_0, D_0, \lambda_0)&=e^{(2\chi + z + d)l}C_0(e^l \bm k, e^{zl}w, \gamma(l), D(l), \lambda(l))\\
    &=e^{(2\chi + z + d)l}\dfrac{2D(\log(\Lambda/|\bm k|))|\bm k|^2e^{2l}}{w^2e^{2zl}+\gamma^2(\log(\Lambda/|\bm k|))|\bm k|^6e^{6l}}\\
    &=\dfrac{2D(\log(\Lambda/|\bm k|))|\bm k|^2}{w^2+\gamma^2(\log(\Lambda/|\bm k|))|\bm k|^6}\\
    &\simeq \dfrac{2D_0\sqrt{2g_0\log(\Lambda/|\bm k|)}|\bm k|^2}{w^2+\gamma_0^22g_0\log(\Lambda/|\bm k|)|\bm k|^6}.
    \end{split}
    \label{eq: correlation function running divergence}
\end{equation}
Where we again used $z=3$ and $\chi=(1-d)/2$ and the last line follows in the limit $|\bm k|\ll \Lambda$. We can also find the static correlation function in the low $\bm k$ limit:
\begin{equation}
    \begin{split}
    S_h(\bm k)&=\int_{-\infty}^{\infty}C(\bm k, w)\dfrac{dw}{2\pi}=\dfrac{D_0}{\gamma_0|\bm k|},
    \label{eq: unchanged structure factor}
    \end{split}
\end{equation}
which is the same as the one obtained in the linear regime. From Eq.~\eqref{eq: correlation function running divergence}, we obtain that the characteristic timescale of relaxation of a mode $\bm k$ is: $\tilde \tau_{\rm ss}\sim |\bm k|^{-3}/\sqrt{\log(\Lambda/|\bm k|)}$ different from the linear result: $\tau_{\rm ss}\sim |\bm k|^{-3}$. We can equivalently find: 
\begin{equation}
    W^2(L\to\infty, t\gg1/(\Lambda^3\gamma))\sim \log\left(t\sqrt{\log(t)}\right)/3 \quad \text{and}\quad \tau_{\rm ss}=\dfrac{L^3}{\sqrt{\log( L^3)}}.
\end{equation}
Therefore, in $d=1$, the relaxation time acquires a logarithmic correction compared to the linear case while the roughness of the interface is unaffected by the non-linearity. The measurement of the relaxation time difference is however difficult.

\end{document}